\documentclass{article}

\usepackage{PRIMEarxiv}

\usepackage[utf8]{inputenc}
\usepackage[T1]{fontenc}
\usepackage[numbers]{natbib} 
\usepackage{hyperref}
\usepackage{url}
\usepackage{booktabs}
\usepackage{amsfonts}
\usepackage{nicefrac}
\usepackage{microtype}
\usepackage{graphicx}
\graphicspath{{media/}}
\usepackage{float}

\usepackage{wrapfig}
\usepackage{fancyhdr}
\title{When Purple Perceived Only at Fixation: 

A Fixation and Distance-Dependent Color Illusion
\thanks{\textit{\underline{Citation}}: \textbf{Schulz-Hildebrandt, H. (2025). \textit{When Purple Perceived Only at Fixation: A Fixation- and Distance-Dependent Color Illusion.}[Preprint]. arXiv. 
\href{https://doi.org/10.48550/arXiv.2509.11582}{https://doi.org/10.48550/arXiv.2509.11582
}}}}

\author{
  Hinnerk Schulz-Hildebrandt\textsuperscript{1}\\
  \textsuperscript{1}Wellman Center for Photomedicine, Massachusetts General Hospital,
  Harvard Medical School, Boston, MA, United States\\
  \texttt{hschulz-hildebrandt@mgh.harvard.edu}
}

\begin{document}
\maketitle

\begin{abstract}
In this paper, an optical illusion is described in which purple structures are perceived as purple at the point of fixation, while surrounding structures of the same purple color are perceived toward a blue hue. As the viewing distance increases, a greater number of purple structures revert to a purple appearance.
\end{abstract}


\section{Introduction}

The perception of color is not a direct consequence of the wavelength of electromagnetic radiation but an active construction of the human visual system, including the involvement of color receptors in the retina and the visual cortex. \citep{RN280} In the human eye, there are three different types of color receptors called cones. The S-cones are stimulated by light with a short wavelength (blue), the M-cones are stimulated by light with a medium wavelength (green), and the L-cones are stimulated by long-wavelength light (red). Although many colors such as blue, green, yellow, orange and red can be assigned to a specific spectral wavelength, \citep{RN283} purple does not exist as a single spectral wavelength within the visible spectrum. It is a non-spectral color that is generated when L- and S-cones are stimulated by the absence of significant signal from M-cones. \citep{RN284} This makes purple a fragile and unstable perception that is easily influenced by physiological and contextual factors \citep{RN291}.

Beyond neural interpretation, the distribution of different cones within the retina also defines the visual appearance of purple. In the fovea, the area of sharpest vision, L and M cones are present in high density, enabling the finest detail and color distinctions. The S-cones make up only 8-12\% of all cones in the retina. \citep{RN286} and are almost completely absent in the absolute center of the fovea.\citep{RN287} Furthermore the macula pigment within the retina exhibits strong absorption in the short-wavelength region of the visible spectrum with a maximum density in the center of the fovea that gradually decreases toward the periphery. \citep{GARDASEVIC201972,kitaoka2025macula} Although the combination of L and S signals is crucial for the perception of purple, the visual system in the fovea must reconstruct purple primarily from the strong red signals of the L-cones and the marginal signals of the S-cones.\citep{RN290} In addition to these biological and optical factors, the perception of colors is not absolute and isolated, but depends on the context and their integration into the ambient scene.\citep{RN288}

The combination of these three mechanisms, the spatial variation of the sensitivity for blue light, the constructive nature of purple and classic color contrast effects, leads to a unique visual illusion. 

\section*{Methods}
\begin{figure}[h]
\centering
\includegraphics[width=0.65\linewidth]{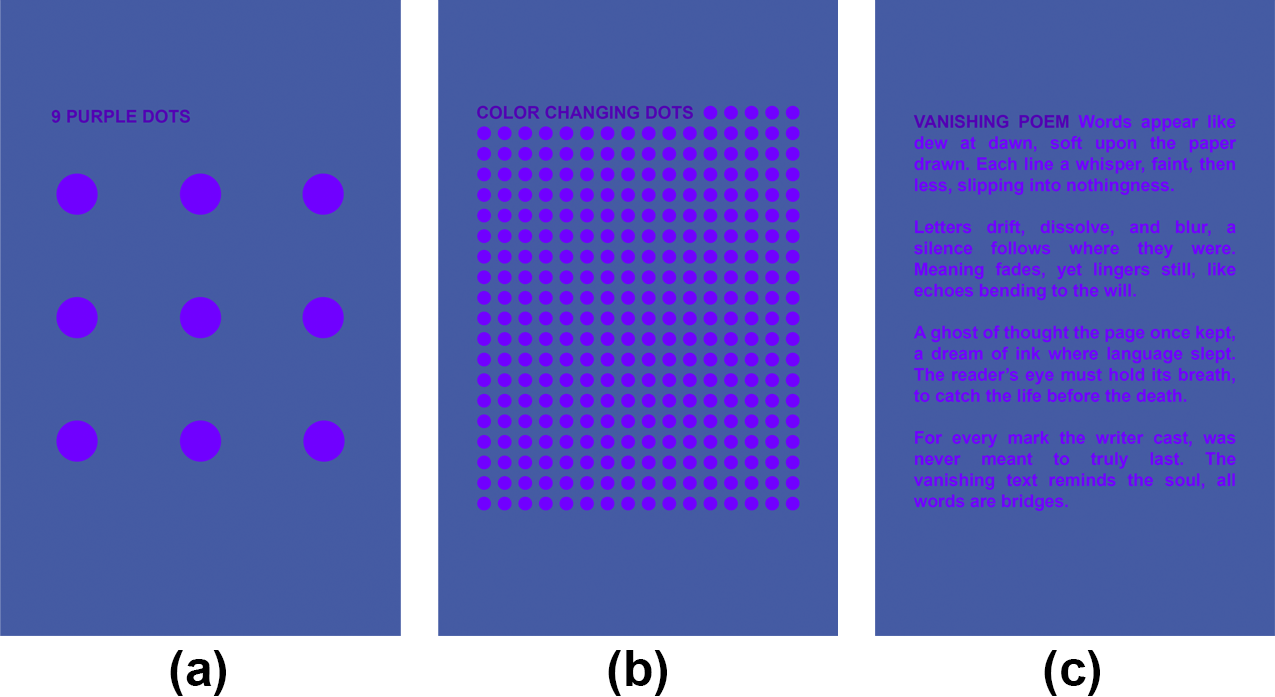}
    \caption{True color representation of the three different optical illusions:
    (a) the ``Purple Dots'', (b) the ``Color Changing Dots'' and (c) the ``Vanishing Poem''.}
    \label{fig:3}
\end{figure}
To demonstrate this effect, a pattern of 9 purple dots [Fig.\,\ref{fig:3}(a)], a smaller sized dot pattern [Fig.\,\ref{fig:3}(b)], and a written text [Fig.\,\ref{fig:3}(c)] were developed.
The "9 Purple Dots" illusion consists of 9 purple dots arranged in a square pattern. Each dot has a diameter of 150\,pixel\,(px)  and is separated by 300\,pixel each other. 
The "Color-Changing Dots" illusion consists of 18 x 20 purple dots with a diameter of 50 px and a grid spacing of 25 px. The dots have a purple color tone, consisting of a mixture of red and blue with the absence of green (Hex:\,\#6600FF). 
The "Vanishing Poem" illusion consists of 94 words written in Arial Bold with a font size of 32 pt, the text has the same purple color tone as the dot pattern. The text of the poem was generated using a large language model (chatGPT, OpenAI). For the demonstration here, a blue background color consisting of a mixture of the three color channels (HEX:\,\#495A9E).

\begin{figure}[h]
\centering
\includegraphics[width=0.65\linewidth]{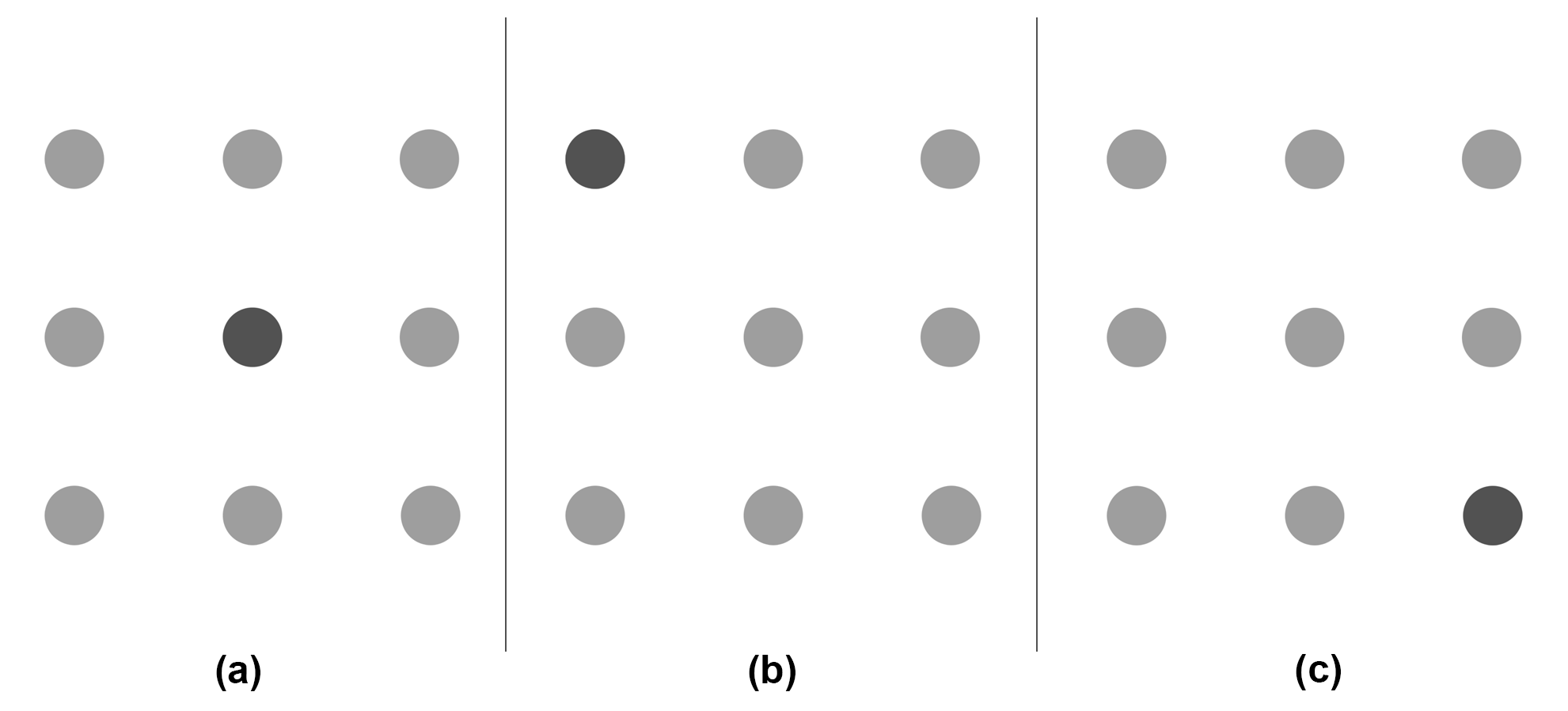}
\caption{Visual perception of the "9 Purple Dots" illusion. Only the fixated dot appears purple; surrounding dots look bluish. Shifting gaze moves the purple perception.}
\label{fig:4}
\end{figure}
When holding the cellphone display at a distance of around 30 cm, the fixed and focused elements were perceived as stable purple. The peripheral elements, on the other hand, appeared bluish, as if the red component of the purple was suppressed. Even when the glaze moves fast over the dot pattern, the purple perception follows the eye movement in real time. Figure \ref{fig:4} shows a schematic representation of the "9 Purple Dots" illusion, when the gaze is fixed at different dots [Fig.\,\ref{fig:4}(a-c)]. This shows that even if the color of the elements and background are not changing, the perception of purple elements is fixation dependent.
\begin{figure}[h]
\centering
\includegraphics[width=0.65\linewidth]{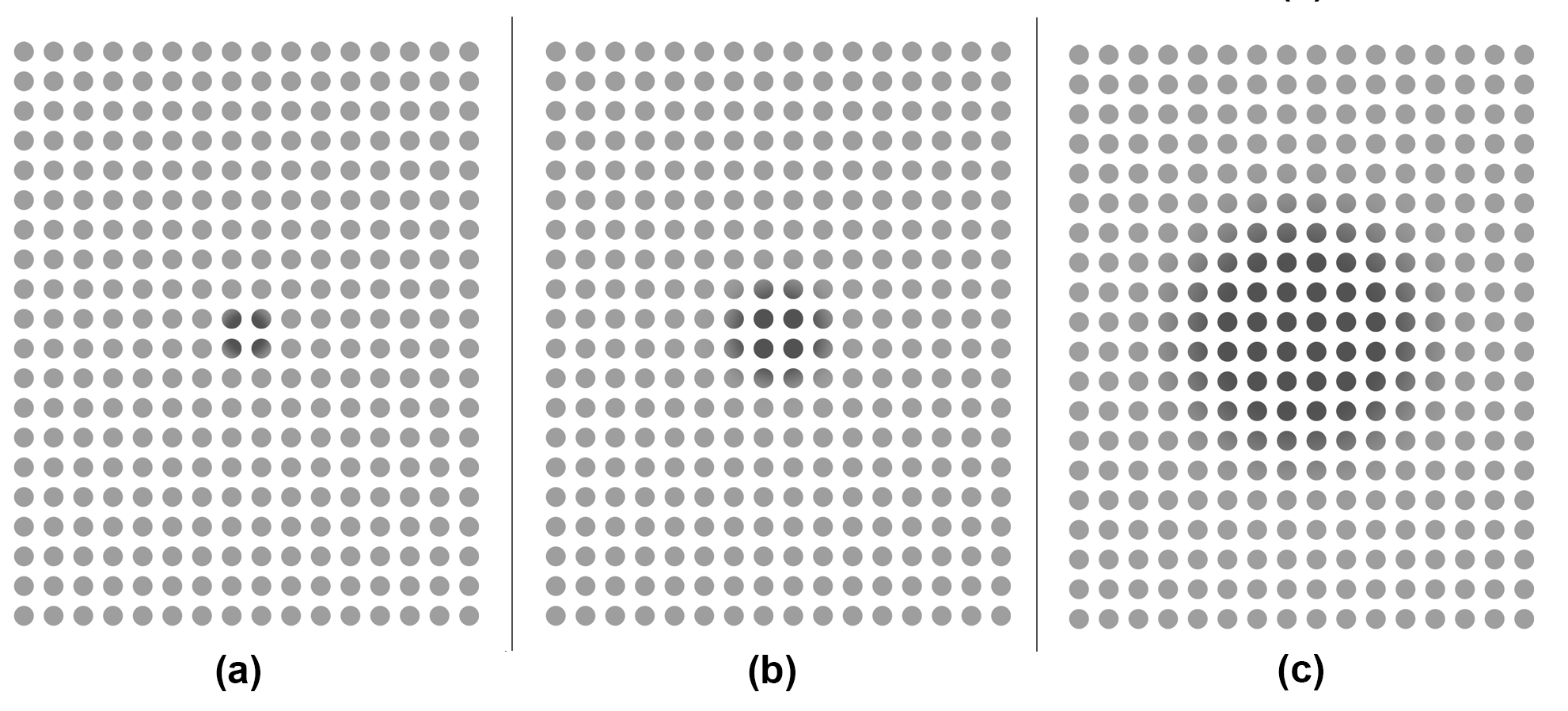}
\caption{Visual perception of the "Color Changing Dots" illusion. At close range few dots look purple; with distance more appear purple until almost all do.}
\label{fig:5}
\end{figure}
Despite the fixation dependency, a distance-dependent effect can also be observed, demonstrated using the "Color Changing Dots" illusion. When the dots are placed close to the eye, at a distance of 10 cm [Fig.\,\ref{fig:5}(a)], only a very small number of dots, appear partially purple. However, as the distance increases, more dots are perceived as purple. At a distance of approximately 30 cm [Fig.\,\ref{fig:5}(b)], roughly twice as many dots are perceived as purple as observed by the author. Moving further away to a distance of 60 cm [Fig.\,\ref{fig:5}(c)] and beyond, the number of dots perceived as purple increases until almost all dots appear purple.
\begin{figure}[h!]
\centering
\includegraphics[width=0.65\linewidth]{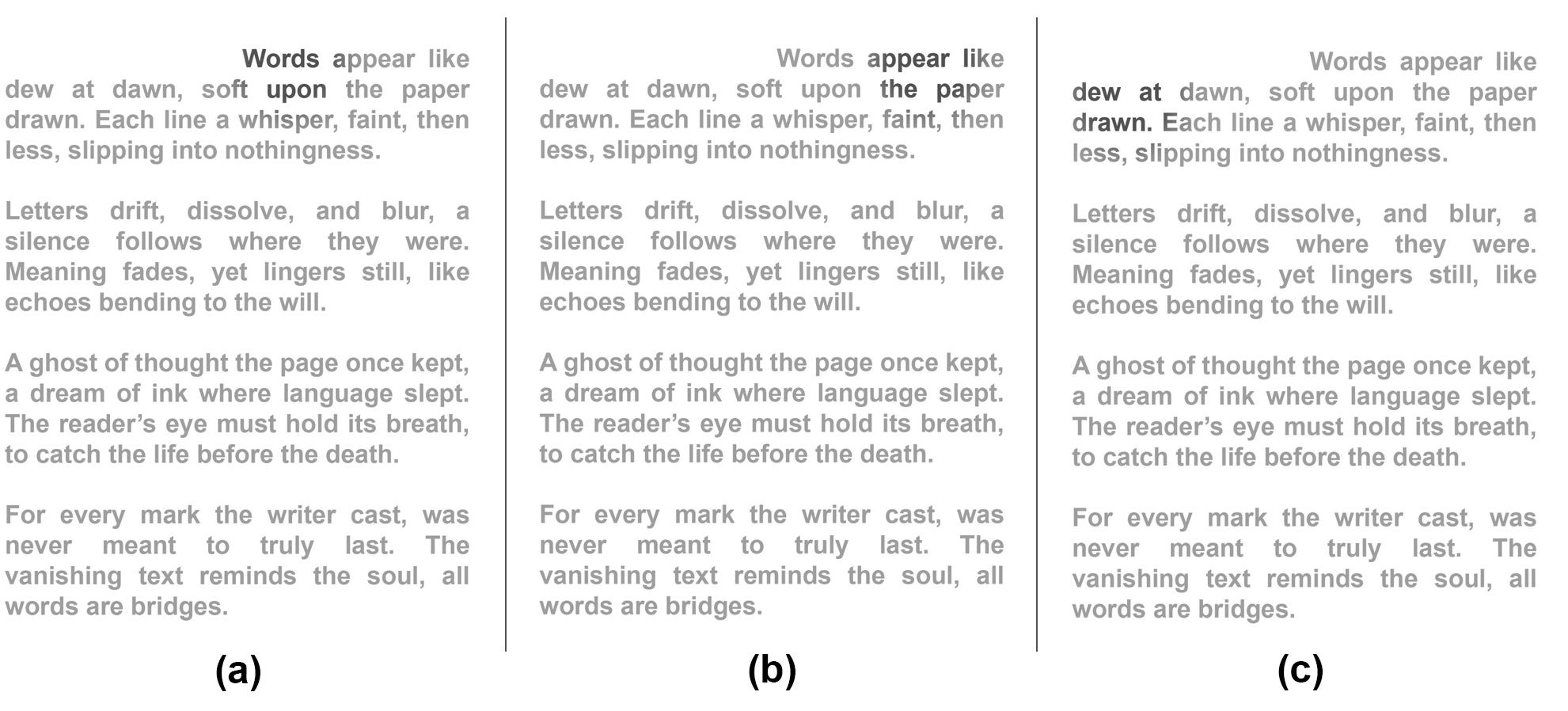}
\caption{Visual perception of the "Vanishing Poem" illusion. Only the word at fixation appears purple; peripheral words shift to bluish.}
\label{fig:6}
\end{figure} Another striking effect visualization of this phenomenon is the "Vanishing Poem" illusion [Fig. \ref{fig:6}]. When holding the smartphone at a comfortable reading distance of around 30 cm, only the fixated words appear purple, while peripheral text shifts toward blue, creating the impression that the text highlights itself during reading.

\section*{Discussion}
A key factor affecting the perceptibility of the effect is the display used. The illusions were initially visualized on an iPhone 16 Pro with an OLED display (2796 × 1290 pixels, 120 Hz). In general, the illusion is most striking on devices with high contrast and a wide color spectrum, but decreases, when the color temperature of the display is shifted towards warmer tones.

Although the origin of the optical illusion was not studied empirically, on the basis of current observations, it is assumed that it is a combination of three factors. First, a classic color contrast effect, which would explain why the periphery structures show up in blueish, second, the reception of purple as a non-spectral color that is created exclusively through the integration of L and S signals, and third, the biological distribution of the cones and absorption of blue light by the macula pigment.

Alterations in color and brightness perception related to the S-cone–free zone and changes in brightness due to wavelength-dependent absorption in the macula have been previously described, particularly in the context of contrast effects. Recently, Kitaoka introduced Maxwell's Spot Illusion, showing a reduction in the brightness of blue dots within central vision, consistent with the optical density profile of macula pigment.\citep{kitaoka2025macula} However, to my knowledge, the specific illusion reported here, which can induce a hue shift for observers, has not previously been described or widely recognized within the vision-science community.

\section*{Acknowledgments}
I thank the vision science community for their valuable feedback and insightful discussions, which have greatly contributed to refining the understanding of the origin of this illusion.

\bibliographystyle{unsrt}
\bibliography{references}

\end{document}